\documentclass{article}%
\usepackage{amsfonts}
\usepackage{amsmath}
\usepackage{amssymb}
\usepackage{charter}
\usepackage{graphicx}%
\providecommand{\U}[1]{\protect \rule{.1in}{.1in}}

\begin{document}

\title{Comment on \textquotedblleft Single-mode excited entangled coherent
states\textquotedblright}
\author{Hong-chun Yuan$^{1}$and Li-yun Hu$^{1,2}$\thanks{Corresponding author:
hlyun2008@126.com}\\$^{1}${\small Department of Physics, Shanghai Jiao Tong University, Shanghai
200030, China}\\$^{2}$ {\small College of Physics and Communication Electronics, Jiangxi
Normal University, Nanchang 330022, China}}
\maketitle

\begin{abstract}
In Xu and Kuang (\textit{J. Phys. A: Math. Gen.} 39 (2006) L191), the authors
claim that, for single-mode excited entangled coherent states $\left \vert
\Psi_{\pm}(\alpha,m)\right \rangle $, \textquotedblleft the photon excitations
lead to the decrease of the concurrence in the strong field regime of
$\left \vert \alpha \right \vert ^{2}$ and the concurrence tends to zero when
$\left \vert \alpha \right \vert ^{2}\rightarrow \infty$". This is wrong.

\end{abstract}

PACS number: 03.65.Ud; 03.67.Hk

In the recent paper \cite{r1}, single-mode excited entangled coherent states
(SMEECSs) $\left \vert \Psi_{\pm}(\alpha,m)\right \rangle $ are introduced and
their entanglement characteristics and the influence of photon excitations on
quantum entanglement are also investigated. They claim that \textquotedblleft
the photon excitations lead to the decrease of the concurrence in the strong
field regime of $\left \vert \alpha \right \vert ^{2}$ and the concurrence tends
to zero when $\left \vert \alpha \right \vert ^{2}\rightarrow \infty$".
Unfortunately, however, this conclusion is wrong.

First, we recall the entangled coherent states (ECS) \cite{r2}%
\begin{equation}
\left \vert \Psi_{\pm}(\alpha,0)\right \rangle =N_{\pm}(\alpha,0)(\left \vert
\alpha,\alpha \right \rangle \pm \left \vert -\alpha,-\alpha \right \rangle ),
\label{01}%
\end{equation}
where $\left \vert \alpha,\alpha \right \rangle \equiv \left \vert \alpha
\right \rangle _{a}\otimes \left \vert \alpha \right \rangle _{b}$ with $\left \vert
\alpha \right \rangle _{a}$ and $\left \vert \alpha \right \rangle _{b}$ being the
usual coherent state in $a$ and $b$ modes, respectively, and
\begin{equation}
\left(  N_{\pm}(\alpha,0)\right)  ^{-2}=2\left[  1\pm \exp \left(  -4\left \vert
\alpha \right \vert ^{2}\right)  \right]  , \label{02}%
\end{equation}
is the normalization constants.

The SMEECSs are obtained throught actions of a creation operator of a
single-mode optical field on the ECSs, which are expressed as%
\begin{equation}
\left \vert \Psi_{\pm}(\alpha,m)\right \rangle =N_{\pm}(\alpha,m)a^{\dagger
m}(\left \vert \alpha,\alpha \right \rangle \pm \left \vert -\alpha,-\alpha
\right \rangle ), \label{03}%
\end{equation}
where without any loss of generality we consider $m$-photon excitations of the
mode $a$ in the ECS and $N_{\pm}(\alpha,m)$ represents the normalization
factor. Using the identity of operator \cite{r3}
\begin{equation}
a^{n}a^{\dagger m}=\left(  -i\right)  ^{n+m}\colon H_{m,n}\left(  ia^{\dagger
},ia\right)  \colon \label{030}%
\end{equation}
where the symbol $\colon \colon$represents the normal ordering for Bosonic
operators $\left(  a^{\dagger},a\right)  $, and $H_{m,n}(\eta,\eta^{\ast})$ is
the two-variable Hermite polynomial\cite{r4},
\begin{equation}
H_{m,n}(\eta,\eta^{\ast})=\sum_{l=0}^{\min(m,n)}\frac{\left(  -1\right)
^{l}n!m!}{l!\left(  m-l\right)  !\left(  n-l\right)  !}\eta^{m-l}\eta^{\ast
n-l}, \label{032}%
\end{equation}
we can easily obtain%
\begin{equation}
\left \langle \alpha \right \vert a^{m}a^{\dagger m}\left \vert \alpha
\right \rangle =m!L_{m}(-\left \vert \alpha \right \vert ^{2}),\text{
\ }\left \langle \alpha \right \vert a^{m}a^{\dagger m}\left \vert -\alpha
\right \rangle =m!e^{-2\left \vert \alpha \right \vert ^{2}}L_{m}\left(
\left \vert \alpha \right \vert ^{2}\right)  , \label{06}%
\end{equation}
and directly calculate the normalization factor
\begin{equation}
\left[  N_{\pm}(\alpha,m)\right]  ^{-2}=2m!\left[  L_{m}\left(  -\left \vert
\alpha \right \vert ^{2}\right)  \pm e^{-4\left \vert \alpha \right \vert ^{2}%
}L_{m}\left(  \left \vert \alpha \right \vert ^{2}\right)  \right]  , \label{07}%
\end{equation}
where $L_{m}(x)$ is the $m-$order Laguerre polynomial defined by\cite{r5}
\begin{equation}
L_{m}(x)=\sum_{l=0}^{m}\frac{(-1)^{l}m!x^{l}}{\left(  l!\right)  ^{2}(m-l)!}.
\label{08}%
\end{equation}
\qquad It is quite clear that when $m=0$, Eq.(\ref{03}) reduces to the usual
ECSs in Eq.(\ref{01}). Eq.(\ref{07}) is valid for any integer $m$ (including
the case of $m=0$) which is different from Eq.(4) of Ref.\cite{r1}.

Next, we calculate the concurrence for the SMEECSs. Noticing that the
photon-added coherent states (PACSs) $\left \vert \alpha,m\right \rangle $ is
defined by\cite{r5}%
\begin{equation}
\left \vert \alpha,m\right \rangle =\frac{a^{\dagger m}\left \vert \alpha
\right \rangle }{\sqrt{m!L_{m}\left(  -\left \vert \alpha \right \vert
^{2}\right)  }}, \label{09}%
\end{equation}
thus the SMEECSs $\left \vert \Psi_{\pm}(\alpha,m)\right \rangle $ in terms of
the PACSs can be rewritten as%
\begin{equation}
\left \vert \Psi_{\pm}(\alpha,m)\right \rangle =M_{\pm}(\alpha,m)(\left \vert
\alpha,m\right \rangle \otimes \left \vert \alpha \right \rangle \pm \left \vert
-\alpha,m\right \rangle \otimes \left \vert -\alpha \right \rangle ), \label{010}%
\end{equation}
where the normalization constant $M_{\pm}(\alpha,m)$ is determined by%
\begin{equation}
\left[  M_{\pm}(\alpha,m)\right]  ^{2}=\frac{L_{m}\left(  -\left \vert
\alpha \right \vert ^{2}\right)  }{2\left[  L_{m}\left(  -\left \vert
\alpha \right \vert ^{2}\right)  \pm e^{-4\left \vert \alpha \right \vert ^{2}%
}L_{m}\left(  \left \vert \alpha \right \vert ^{2}\right)  \right]  }.
\label{011}%
\end{equation}

Following the approach of Ref.\cite{r6} and considering Eqs.(\ref{06}) and
(\ref{09}), for the SMEECSs $\left \vert \Psi_{\pm}(\alpha,m)\right \rangle ,$
the concurrence can be calculated as
\begin{equation}
C_{\pm}(\alpha,m)=\frac{\sqrt{\left(  1-p_{1}^{2}\right)  \left(  1-p_{2}%
^{2}\right)  }}{1\pm p_{1}p_{2}}, \label{011a}%
\end{equation}
where
\begin{equation}
p_{1}=\left \langle \alpha,m\right \vert \left.  -\alpha,m\right \rangle
=\frac{\exp(-2\left \vert \alpha \right \vert ^{2})L_{m}\left(  \left \vert
\alpha \right \vert ^{2}\right)  }{L_{m}\left(  -\left \vert \alpha \right \vert
^{2}\right)  }, \label{012}%
\end{equation}
and \bigskip%
\begin{equation}
p_{2}=\left \langle \alpha \right \vert \left.  -\alpha \right \rangle
=\exp(-2\left \vert \alpha \right \vert ^{2}). \label{013}%
\end{equation}
Then submitting Eqs.(\ref{012}) and (\ref{013}) into Eq.(\ref{011a}) we see
that
\begin{equation}
C_{\pm}(\alpha,m)=\frac{\left[  \left(  L_{m}^{2}\left(  -\left \vert
\alpha \right \vert ^{2}\right)  -e^{-4\left \vert \alpha \right \vert ^{2}}%
L_{m}^{2}\left(  \left \vert \alpha \right \vert ^{2}\right)  \right)  \left(
1-e^{-4\left \vert \alpha \right \vert ^{2}}\right)  \right]  ^{1/2}}%
{L_{m}\left(  -\left \vert \alpha \right \vert ^{2}\right)  \pm e^{-4\left \vert
\alpha \right \vert ^{2}}L_{m}\left(  \left \vert \alpha \right \vert ^{2}\right)
}, \label{014}%
\end{equation}
which is another expression different from Eqs.(23) and (24) in Ref.\cite{r1}.
In particular, when $m=0$, Eq.(\ref{014}) becomes
\begin{equation}
C_{+}(\alpha,0)=\frac{1-e^{-4\left \vert \alpha \right \vert ^{2}}}%
{1+e^{-4\left \vert \alpha \right \vert ^{2}}},C_{-}(\alpha,0)=1. \label{26}%
\end{equation}
It implies that the concurrence $C_{+}(\alpha,0)$ of ECS $\left \vert \Psi
_{+}(\alpha,0)\right \rangle $ increases with the values of $\left \vert
\alpha \right \vert ^{2};$ while $C_{-}(\alpha,0)$ is indenpent of $\left \vert
\alpha \right \vert ^{2}\ $and $\left \vert \Psi_{-}(\alpha,0)\right \rangle $ is
a maximally entangled state.

In order to see clearly the influence of the concurrence with parameter $m$,
the concurrences $C$ for the state $\left \vert \Psi_{\pm}(\alpha
,m)\right \rangle $ as a function of $\left \vert \alpha \right \vert ^{2}$ are
shown in Figs.1 and 2. It is shown that $C_{\pm}(\alpha,m)$ increases with the
increase of $\left \vert \alpha \right \vert ^{2}$ for given parameter $m$.
Especially, the concurrence $C_{\pm}(\alpha,m)$ tend to unit for the larger
$\left \vert \alpha \right \vert ^{2}.$These conclusions are completely different
from those of Ref.\cite{r1}.\

\bigskip \newpage

\begin{figure}[ptb]
\label{fig1}
\centering \includegraphics[width=10cm]{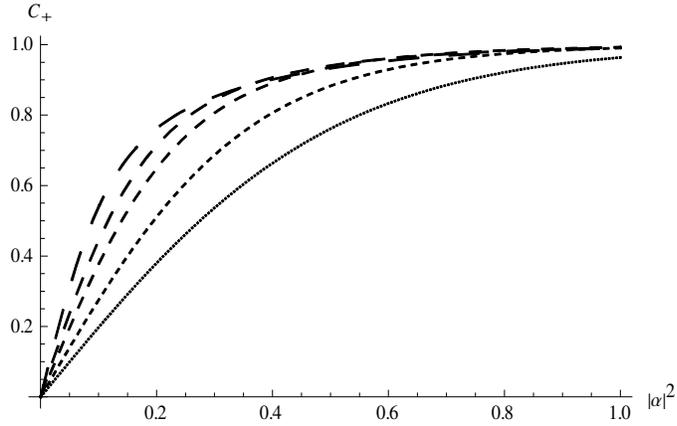}\caption{Concurrence
of entanglement of $\left \vert \phi_{+}(\alpha,m)\right \rangle $ as a function
of $\left \vert \alpha \right \vert ^{2}$ for the different photon excitations
with $m$ $=0$(solid line), $m=1$(dashed line), $m=3$(dotted line),
$m=5$(dash-dotted line), and $m=20$(dash-dash-dotted line), respectively.}%
\end{figure}\begin{figure}[ptb]
\label{fig2}
\centering \includegraphics[width=10cm]{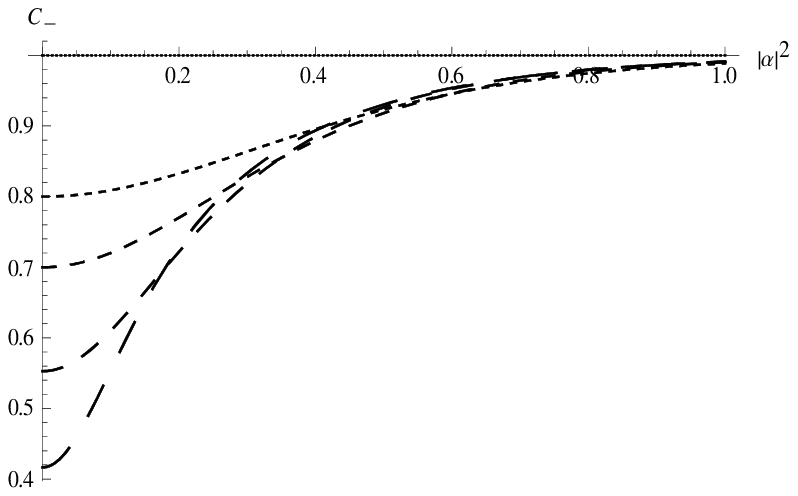}\caption{Concurrence
of entanglement of $\left \vert \phi_{-}(\alpha,m)\right \rangle $ as a function
of $\left \vert \alpha \right \vert ^{2}$ for the different photon excitations
with $m=0$(solid line), $m=3$(dashed line), $m=5$(dotted line), $m=10$%
(dash-dotted line), and $m=20$(dash-dash-dotted line), respectively.}%
\end{figure}

\end{document}